\documentclass[aps,prb,preprint,groupedaddress]{revtex4-1}

\usepackage{graphicx}
\usepackage{dcolumn}
\usepackage{bm}

\begin{document}


\title{Realization of Ordered Magnetic Skyrmions in Thin Films at Ambient Conditions}

\author{Ryan D. Desautels}
\altaffiliation{These two authors contributed equally}
\affiliation{Neutron Scattering Division, Oak Ridge National Laboratory, Oak Ridge, Tennessee 37831, USA}

\author{Lisa DeBeer-Schmitt}
\altaffiliation{These two authors contributed equally}
\email{debeerschmlm@ornl.gov}
\affiliation{Neutron Scattering Division, Oak Ridge National Laboratory, Oak Ridge, Tennessee 37831, USA}

\author{Sergio Montoya}
\affiliation{Space and Naval Warfare Systems Center Pacific, San Diego, CA 92152, USA}

\author{Julie A. Borchers}
\affiliation{NIST Center for Neutron Research, National Institute of Standards and Technology, Gaithersburg, MD 20899, USA}

\author{Soong-Geun Je}
\affiliation{Center for X-ray Optics, Lawrence Berkeley National Laboratory, Berkeley, CA 94270, USA}

\author{Nan Tang}
\affiliation{Department of Materials Science and Engineering, University of Tennessee, Knoxville, TN 37996, USA}

\author{Mi-Young Im}
\affiliation{Center for X-ray Optics, Lawrence Berkeley National Laboratory, Berkeley, CA 94270, USA, 
and School of Materials Science and Engineering, Ulsan National Institute of Science and Technology, Ulsan, Republic of  Korea}

\author{Michael R. Fitzsimmons}
\affiliation{Neutron Scattering Division, Oak Ridge National Laboratory, Oak Ridge, Tennessee 37831, USA and Department of Physics and Astronomy, University of Tennessee, Knoxville TN 37996, USA}

\author{Eric E. Fullerton}
\affiliation{Center for Memory and Recording Research, University of California, San Diego, La Jolla, CA 92093, USA}

\author{Dustin A. Gilbert}
\affiliation{NIST Center for Neutron Research, National Institute of Standards and Technology, Gaithersburg, MD 20899, USA and Department of Materials Science and Engineering, University of Tennessee, Knoxville, TN 37996, USA}

\date{\today}

\begin{abstract}
Magnetic skyrmions present interesting physics due to their topological nature and hold significant promise for future information technologies. A key barrier to realizing skyrmion devices has been stabilizing these spin structures under ambient conditions. In this manuscript, we exploit the tunable magnetic properties of amorphous Fe/Gd mulitlayers to realize skyrmion lattices which are stable over a large temperature and magnetic field parameter space, including room temperature and zero magnetic field. These hybrid skyrmions have both Bloch-type and N\'eel-type character and are stabilized by dipolar interactions rather than Dzyaloshinskii-Moriya interactions, which are typically considered required for the generation of skyrmions. Small angle neutron scattering (SANS) was used in combination with soft X-ray microscopy to provide a unique, multi-scale probe of the local and long-range order of these structures. These results identify a pathway to engineer controllable skyrmion phases in thin film geometries which are stable at ambient conditions.
\end{abstract}

\maketitle

\section*{Introduction}

Skyrmions have emerged as one of the most promising approaches to realize next-generation, ultra-low power memory and logic devices\cite{Fert.2013,Woo.2016,Zhang.2015,Jiang.2017}. The small spatial size of these magnetic structures (down to the near atomic level\cite{Heinze.2011}) coupled with their relative ease of current induced mobility\cite{Jonietz.2010} and topologically enhanced stability,\cite{Fert.2013} make these chiral spin textures among the most exciting emerging spintronic technologies. The ability to stabilize skyrmion structures in environments relevant to consumer technologies remains a challenging barrier to realizing skyrmion-based devices today. Specifically, until recently, skyrmions were only ever demonstrated in temperature windows well below room temperature,\cite{Nagaosa.2013, Jiang.2017} and always in finite magnetic fields\cite{Karube.2017, Tokunaga.2015}. Advances, particularly in interfacial Dzyaloshinskii-Moriya (DM) interactions,\cite{Moreau.2016} have largely overcome the challenges associated with room-temperature stability.\cite{Jiang.2017, Jiang.2016} Few approaches have been demonstrated that can stabilize skyrmions at zero applied magnetic field, but include rapid quenching,\cite{Karube.2016,Karube.2017} pulsed electrical currents,\cite{He.2017} geometric confinement,\cite{Zheng.2017, Boulle.2016} or pinning the skyrmions with nanostructures\cite{Sun.2013, Gilbert.2015}. These approaches place restrictions on the skyrmion system, limiting both fundamental research and eventual device architectures. Removing the requisite magnetic field, especially at room temperature, in systems relevant to eventual skyrmion devices is crucial to integrating skyrmions into spintronic architectures\cite{Dieny.1991}.

We demonstrate the realization of an ordered magnetic skyrmion lattice in amorphous multilayer thin-films without DM interactions. These skyrmions have been suggested to be stabilized by dipole-dipole interactions,\cite{Montoya.2017-1,Montoya.2017-2} but recent works have suggested a random anisotropy (RA) specific to amorphous and nanocrystalline systems may also contribute to skyrmion formation\cite{Chudnovsky_2018}. Utilizing a straight forward magnetic field sequence (presented in Supplemental Fig.~S1), the labyrinthine domains typically present at remanence are ordered into artificial stripe domains, analogous to traditional B20-structured skyrmion materials. The skyrmion regime is accessed by increasing the magnetic field from the stripe state, breaking up the stripes and precipitating arrays of long-range ordered skyrmions. Once generated, these dipole-stabilized skyrmions\cite{Montoya.2017-1,Montoya.2017-2} are stable over a field range from  200~mT to -55~mT, including zero magnetic field, and a temperature range from 10~K to 320~K, demonstrating good alignment with the parameter space of consumer technologies.

Unlike traditional skyrmion systems which are stabilized by DM interactions, these dipole-stabilized skyrmions possess no symmetry breaking in the in-plane circularity and thus both chiralities can be realized.  However, this is not the case at the surfaces of the skyrmion columns. Specifically, these dipole-stabilized skyrmions possess flux-closure domains at the top and bottom surfaces, which have a structure analogous to N\'eel-type skyrmions. The chirality of these N\'eel-type surfaces ($\textit{i.e.}$ caps) is determined directly by the dipole field emanating from the core of the skyrmion and is thus uniform for all of the skyrmions with a common core orientation. Further the chirality of the top surface is opposite to the chirality of the closure domains of the bottom surface. Between the N\'eel caps, away from the top and bottom surface of the film, the moments wrap into Bloch-type structures with random chirality developing a hybrid skyrmion structure with an ordered lattice stabilized by the mutual repulsion of these caps (presented in Supplemental Fig. S1c).  The dipole-stabilized skyrmions in our Fe/Gd system open up a new rich playground of physical phenomena as well as to explore for emergent technologies\cite{Jiang.2017}.

\section*{Methods}

The Fe/Gd multilayer films were deposited on 1 cm$^2$ Si substrates by DC magnetron sputtering.  Deposition was performed at room temperature in an ultrahigh vacuum with a 3~mTorr (1~Torr = 133~Pa) argon environment\cite{Montoya.2017-1,Montoya.2017-2}. The multilayer films were grown by depositing alternating layers of Fe and Gd until the desired number of layers was achieved.  In all cases, 5~nm tantalum seed and capping (to prevent oxidation) layers were used.  This is illustrated in Fig. S2a.  A picture of the sample in the magnet bore is shown in Fig. S2b.

Two-dimensional small angle neutron scattering (2D SANS) experiments were performed at the GP-SANS beamline at Oak Ridge National Laboratory's (ORNL) High Flux Isotope Reactor (HFIR)\cite{Heller.2018, Wignall.2012}.  These experiments were performed at room temperature and varying the magnetic field using a 5~T horizontal open bore dry cryogenic superconducting magnet, with the magnetic field aligned parallel to the neutron beam (Supplemental Fig.~S2c). The SANS instrument was configured to use 16~\AA~neutrons ($\Delta\lambda/\lambda$~=~0.13) with a detector distance of 19.2~m on a co-linear aligned stack of multiple (12 for the [(Fe(3.6~\AA)~/~Gd(4.0~\AA)]~$\times120$ and 10 for the [(Fe(3.4~\AA)~/~Gd(3.8~\AA)]~$\times120$) films.  This sample holder ensures that the films remain aligned co-linearly and restricts any in-plane rotation/slipping of the film stack.  An 8~mm diameter aperture was used to fully illuminate the films in the beam. The temperature dependence 2D SANS measurements were performed on the Very Small Angle Neutron Scattering (VSANS) beamline at the National Institute for Standards and Technology (NIST) using 14~\AA~neutrons, a water cooled wire-wrap electromagnet, and a closed-cycle refrigeration system. 

The artificially aligned stripe domains were stabilized by rotating the sample about the x-axis (Supplemental Fig. S1b, out of the figure plane in Supplemental Fig. S2c) by $\sim$45$^{\circ}$ and applying a saturating field of $\mu_{0}H=$~500~mT. Reducing the field from saturation would typically nucleate randomly oriented labyrinth domains, however, by tilting the sample, the in-plane projection of the field breaks the symmetry, and orders the domains as stripes domains along the in-plane field direction (stripes aligned parallel to the field direction). After removing the magnetic field, the sample was rotated to its original configuration, with the field along the film normal, (Supplemental Fig. S2c) for the SANS measurements.

X-ray microscopy images were captured on BL 6.3.2 at the Advanced Light Source. Images were taken using the Fe $L_2$ edge, measuring the transmission with clockwise and counter-clockwise circularly polarized X-rays. Taking the difference between the two polarizations generates the magnetic contrast by the X-ray magnetic dichroism. To achieve transmission, the samples were grown on SiN membrane windows. 

Micromagnetic simulations were performed using the Object Oriented Micromagnetic Framework (OOMMF) using a saturation magnetization of 800~emu/cm$^{3}$ (1~emu/cm$^{3}$ = 10$^{3}$~A/m), Exchange stiffness of 1$\times10^{-11}$~J/m, and uniaxial anisotropy of $200\times10^{3}$~J/m$^{3}$; dipole interactions were included but no DM interaction were included. Stability simulations were performed at $H=0$. The skyrmion size accurately reproduced the experimental results, with diameters of $\approx$200 nm.

\section*{Results and Discussion}

\begin{figure*}[ht]
    \centering
    \includegraphics[scale = 0.15]{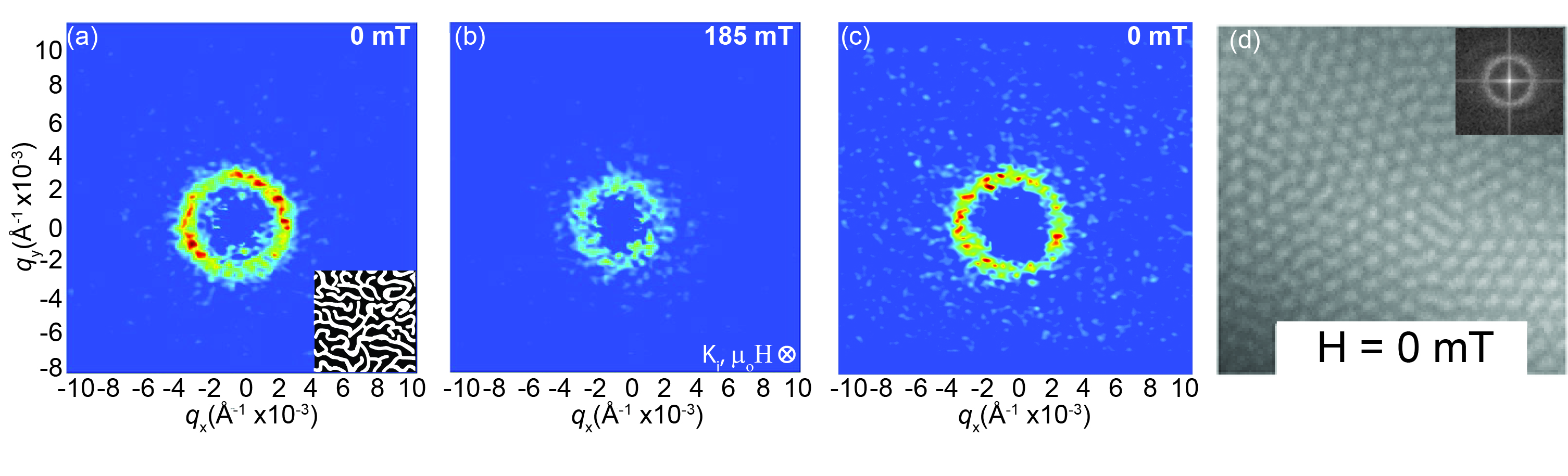}
    \caption{Disordered skyrmions in a continuous amorphous thin film.  Initially, the [(Fe(3.6~\AA)~/~Gd(4.0~\AA)]$\times$120 film was saturated with a field parallel to the incoming neutrons and normal to the films (see Supplemental Fig.~S2a).  SANS patterns collected at; (\textbf{a}) remanance ($\mu_{0}H=$~0~T), (\textbf{b}) within the disordered skyrmion state ($\mu_{0}H=$~185~mT), (\textbf{c}) and back at remanence from the disordered skyrmion state. (\textbf{d}) Soft X-ray microscopy images taken at $\mu_{0}H=$~0~T, after analogous preparation.}
    \label{fig:2DSANS_disorder}
\end{figure*}

Measurements were performed on multilayer thin-films of [(Fe(3.6~\AA)~/~Gd(3.8~\AA)]$\times$120 and [(Fe(3.4~\AA)~/~Gd(4.0~\AA)]$\times$120 oriented perpendicular to the incoming neutron beam and external magnetic field. At remanence, the domains coalesced into long interwoven 'worm-like' structures (labyrinthine) with similar widths (illustrated in the inset of Figure~\ref{fig:2DSANS_disorder}a)\cite{Montoya.2017-1,Montoya.2017-2}. The Fourier transform of these disordered domains defines the small angle neutron scattering (SANS) pattern,\cite{Milde.2013} and was observed to be a ring (Figure~\ref{fig:2DSANS_disorder}a), as expected. When approaching saturation, the disordered labyrinthine domains break apart to form domains of magnetic skyrmions with local hexagonal ordering, but no long range orientation of the hexagonal lattice\cite{Montoya.2017-2}. The existence of ordered skyrmion lattices is not expected in the RA model,\cite{Chudnovsky_2018} but would be present in the dipole model - due to their repulsive interactions - which further suggests that dipole interactions are the dominant factor in these skyrmions. Each of the individual textures resulting from the dissolution of the labyrinthine domains possess a continuous, closed loop boundary and antiparallel core and perimeter, characteristic of a Bloch skyrmion. The integrated solid angle of these structures - defining their topological charge - can indeed be unity, making them homotopically equivalent to skyrmions observed in traditional B20 materials\cite{Nagaosa.2013} or 3d-5d multilayers\cite{Moreau.2016}. One distinction of these skyrmions is that they are not stabilized by the DM interaction. The likely stabilizing mechanism is dipole interactions between the skyrmion core and the surrounding (antiparallel) matrix; the RA mechanism is expected to play a small role here due to the absence of orbital momentum in the gadolinium,\cite{hellman1998specific} which results in minimal magnetocrystalline anisotropy. Neither of these mechanisms have a geometric symmetry-breaking, as the DM interaction does, and thus there is no net chirality in the lattice; the two chiral states are energetically degenerate and thus 'writing' to the chirality for e.g. data storage applications could be achieved. The skyrmions realized by applying the magnetic field strictly along the film normal are form locally ordered domains with no long-range orientation, and the SANS pattern from all the domains collectively contribute to a broad ring feature in the SANS pattern, indicating the approximate equidistant spacing of the skyrmions. Increasing the magnetic field from the labyrinth phase at $\mu_{0}H$~=~0 to $\mu_{0}H$~=~185~mT, forming skyrmions, the SANS pattern evolves as described, Figure~\ref{fig:2DSANS_disorder}b. One may notice that both the labyrinth and skyrmion phases (Figure~\ref{fig:2DSANS_disorder}a-b) form a ring feature, indicating each possesses long-range periodicity but no orientation and are indistinguishable by SANS. Returning the sample to remanence without going through saturation does not affect the ring structure in SANS but using soft X-ray microscopy(Figure~\ref{fig:2DSANS_disorder}d) one can see the skyrmions are present at zero applied field. Calculating the Fourier transform of the X-ray image, Figure~\ref{fig:2DSANS_disorder}d inset, shows a ring - the observed banding is an artifact of the image processing and does not correspond to any experimental feature.  Soft X-ray microscopy provides a local, real-space image to distinguish between the disordered skyrmion state and labyrinth states that look the same with SANS. SANS is an ensemble average and is related to the Fourier transform of the X-ray data. More X-ray images are presented in the Supplemental Materials in Figure S3 which show the field evolution of the skyrmions in the disordered state.

\begin{figure*}[ht]
    \centering
    \includegraphics[scale = 0.15]{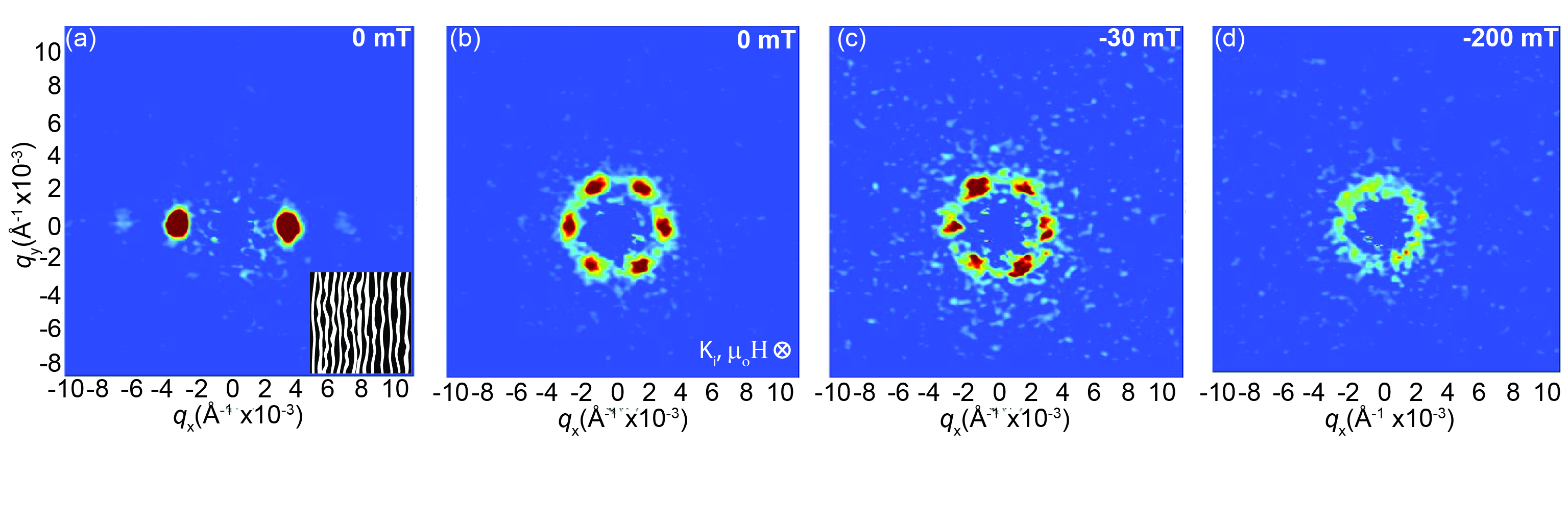}
    \caption{Measured SANS pattern at sequential fields for [(Fe(3.6~\AA)~/~Gd(4.0~\AA)]$\times$120. The sample was then saturated in the rotated geometry (Supplemental Fig.~S1b) then the field reduced to remanence and the films were rotated to their original orientation ,(\textbf{c}) the SANS pattern was measured, showing two peaks corresponding to a well-ordered stripe domain. The out-of-plane magnetic field was increased to $\mu_{0}H=$~185~mT (not shown). SANS patterns were measured as the magnetic field was reduced towards negative saturation, with measurements shown at (\textbf{b}) $\mu_{0}H=$~0~T (skyrmion phase), (\textbf{c}) $\mu_{0}H=$~-30~mT, (\textbf{d}) $\mu_{0}H=$~-200~mT.}
    \label{fig:2DSANS_order}
\end{figure*}

Comparing the skyrmions and labyrinthine domain structure here to traditional skyrmion materials,\cite{Nagaosa.2013} the skyrmion phase in B20 materials is bounded by chiral stripe phases at lower fields and temperatures\cite{Muhlbauer.2009}. We propose that by promoting order within the labyrinth structure, an artificial stripe phase can be realized which mimics the stripe phase in the traditional materials and the resultant skyrmions may form with long-range coherent order. Ordering of the domains was accomplished by first rotating the sample's surface normal by $\approx45^{\circ}$ relative to the magnetic field (Supplemental Fig.~S1b) after which a saturating field (500~mT) was applied, then the field was reduced to remanence. Saturation of the sample in the rotated geometry induces an in-plane field projection which breaks the symmetry and orders the labyrinthine domains into oriented stripe domains, see inset of Figure~\ref{fig:2DSANS_order}a. The sample was then rotated back to its original orientation - with the surface normal parallel to the neutrons and magnetic field - and the SANS pattern was measured. The SANS measurement shows two bright peaks along the axis of the film's rotation (Fig.~\ref{fig:2DSANS_order}a), indicating that the labyrinthine domains are ordered, with stripes forming parallel to the in-plane field projection during the saturation sequence.

Increasing the magnetic field from remanence to $\mu_{0}H=$~185~mT, applied along the film's surface normal, the SANS pattern evolves from two peaks located on the horizontal ($Q_x$) axis (which also was the axis of rotation) to six peaks located at 60$^{\circ}$ increments. Reducing the magnetic field to $\mu_{0}H$~=~0~mT, the hexagonal pattern is unchanged (Figure~\ref{fig:2DSANS_order}b). The 6-fold pattern is widely associated with the skyrmion phase and, compared to the ring pattern in Figure~\ref{fig:2DSANS_disorder}b and c, indicates that the skyrmions form hexagonally packed arrays with long range order and orientation. This ordering is expected to occur by setting one of the nearest neighbor axes of the emergent hexagonal skyrmion lattice with the stripe domain orientation, while the other two axes coordinate to minimize the skyrmion-skyrmion repulsive interaction.

At $\mu_{0}H=0$ the SANS feature appears at $|Q|$~=~0.0025(8)~\AA$^{-1}$, indicating a skyrmion center-to-center separation of 230 nm; this value gives the upper limit to the skyrmion diameter. The lattice's field stability was determined by reducing the out-of-plane magnetic field towards negative saturation (Figure~\ref{fig:2DSANS_order}b-d). As the magnetic field is reduced from remanence ($H=0$) to -30~mT (Fig.~\ref{fig:2DSANS_order}c), two peaks become significantly stronger, indicating the re-emergence of a co-existing stripe domain phase, in agreement with X-ray microscopy images (see Supplemental Fig.~S3). The new stripe phase is rotated by 60$^\circ$ relative to the initial orientation in Figure~\ref{fig:2DSANS_order}c. The difference in orientation may be due to a slight misalignment of the sample with respect the field in the y-axis which would give a small in-plane component in that direction, or subtle texturing within the film. Approaching saturation, the SANS pattern becomes a broad ring of scattering, shown in Figure~\ref{fig:2DSANS_order}d, suggesting a nearly complete loss of orientation. After achieving magnetic saturation, the SANS pattern collapses to $Q=0$, \textit{i.e.} no features were observed in the measurements (not shown). Similar results were observed for the [(Fe(3.6~\AA)~/~Gd(3.8~\AA)]$\times$120 stack of films (Supplemental Fig.~S2).  This further emphasizes the long-range order which is achieved after conditioning in the tilted geometry. These results validate the tilted field geometry conditioning as an effective way to transform the disordered labyrinth domains and resultant sea of skyrmions into an ordered skyrmion lattice. All previous results of these systems were done on disordered skyrmions\cite{Montoya.2017-1,Montoya.2017-2}. This approach presents a simple procedure to manipulate and order skyrmions which, in this system, remain ordered and stable over an extended temperature and field range.

\begin{figure*}[ht]
    \centering
    \includegraphics[scale =0.2]{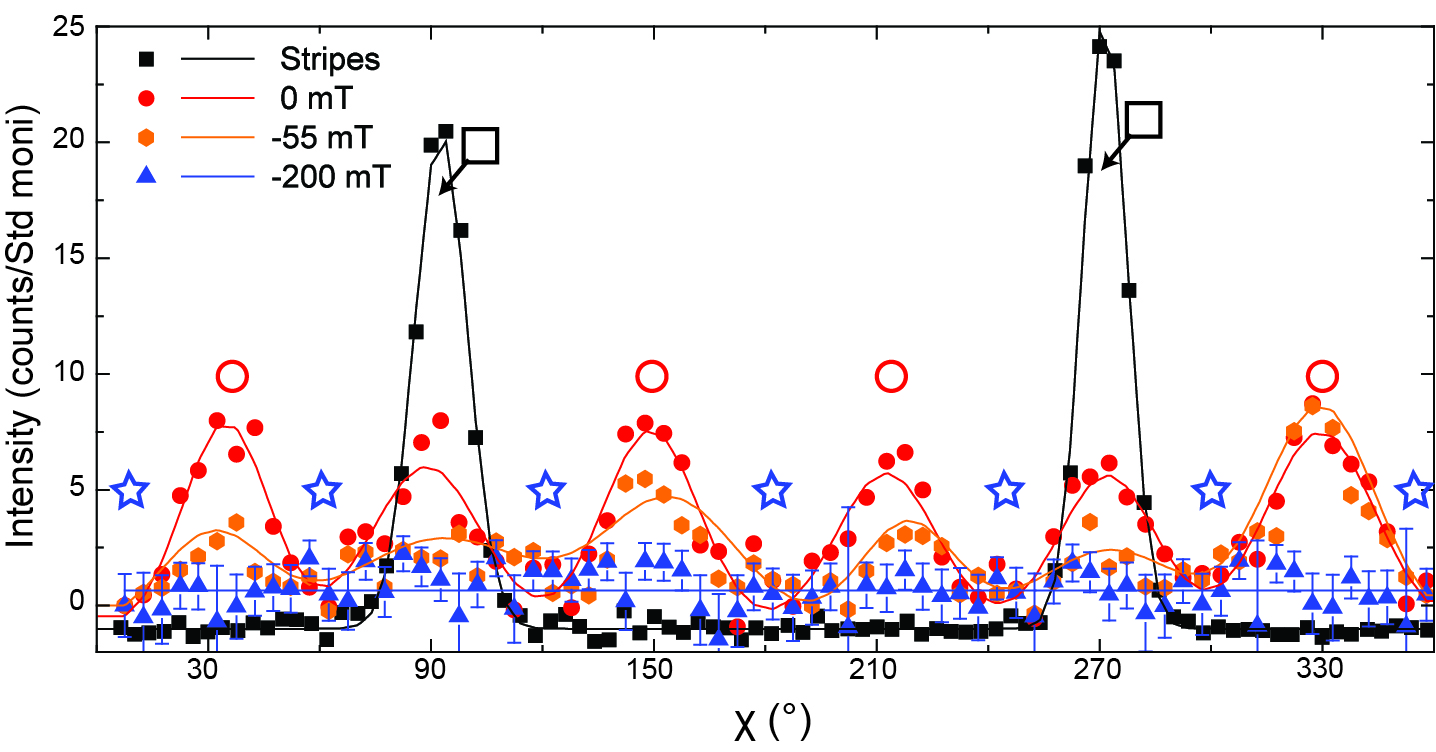}
    \caption{Azimuthal projection of the intensity for a \textit{Q}~=~0.0025~\AA$^{-1}$ for a variety of fields. For $\mu_{0}H=$~0~mT and  (black squares) after saturation in the tilted geometry,the data shows strong scattering intensity at $\chi$~=~90$^{\circ}$ and 270$^{\circ}$, along the axis of rotation. Increasing the magnetic field to $\mu_{0}H=$~185~mT then returning towards negative saturation the azimuthal projection shows a hexagonal pattern with peaks separated by $\approx$~60$^{\circ}$ associated with the skyrmion state, followed by an enhancement of the peaks at 150$^{\circ}$ and 330$^{\circ}$ indicating the coexistence phase, and at large negative values to an azimuthally symmetric ring.  The circle, star, and square shapes represent the position where intensities were measured, as described in the Supplemental Materials.}
    \label{fig:Analysis}
\end{figure*}

To further investigate the domain transitions, a projection of the intensity as a function of azimuthal angle, $\chi$ (defined in Supplemental Fig.~S2d), was taken at the peak intensity ($|Q|$~=~0.0025~\AA$^{-1}$). Representative plots for each of the main phases are presented in Figure~\ref{fig:Analysis}. The transition from the artificial striped domain (solid black squares) to the skyrmion lattice phase (solid red circles) is readily identifiable by the change in the azimuthal pattern from two peaks to the six-peaks. We note also that these peaks are separated by 180$^{\circ}$ and 60$^{\circ}$, respectively, identifying the two-fold and regular six-fold patterns in the 2D SANS measurements in Figures~\ref{fig:2DSANS_order}c-d. The coexistence phase (solid orange hexagons), which is highlighted at $\mu_{0}H=$~-30~mT, (Figure~\ref{fig:2DSANS_order}c) has a six-fold symmetry with additional intensity at $\chi$~=~150$^{\circ}$ and 330$^{\circ}$; the increase in intensity at these two peaks is consistent with the re-emergence of stripe domains. Approaching magnetic saturation the coexistence phase loses long-range structure, resulting in a collapse in the SANS features into a broad ring, corresponding to an approximately constant intensity at all angles (solid blue triangles).

\begin{figure}[ht]
    \centering
    \includegraphics[scale = 0.2]{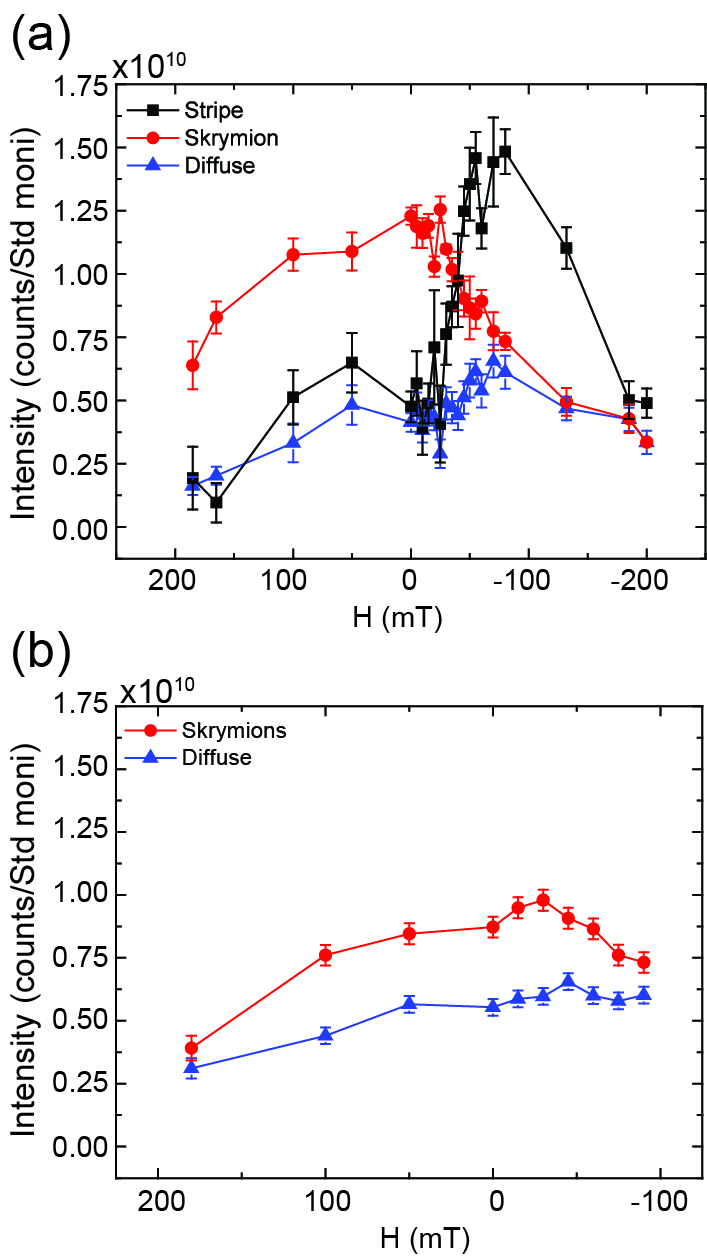}
    \caption{ Scattering intensity from the ring-like background, skyrmion and stripe phases in (\textbf{a}) the [(Fe(3.4~\AA)~/~Gd(4.0~\AA)]$\times$120 films and (\textbf{b}) the [(Fe(3.6~\AA)~/~Gd(3.8~\AA)]$\times$120 films are shown as a function of decreasing magnetic field. No co-existence phase was seen in the [(Fe(3.6~\AA)~/~Gd(3.8~\AA)]$\times$120 sample. Axis progression reflects the progression of the measurement sequence.}
    \label{fig:Analysis2}
\end{figure}

The azimuthal projections show that there are regions in the SANS pattern which can be associated with each of the magnetic configurations. Tracking the intensities of these special regions, identified explicitly as circles, stars, and squares, in Figure~\ref{fig:Analysis} and Supplemental Figure S2d, allows us to follow the transitions between the magnetically ordered phases. The field dependence of the scattering intensity from the identified regions, for the [(Fe(3.4~\AA)/Gd(4.0~\AA)]$\times$120 film stack, is shown in Figure~\ref{fig:Analysis2}a and follows the same field sequence as described above (saturated in a rotated configuration, return to remanence, rotate to a normal geometry, increase the magnetic field to $\mu_{0}H=$~185~mT, then measure under a decreasing magnetic field). The plotted intensity for the ring-like background and skyrmions are as-measured; the intensity of the stripe phase is defined as the measured intensity, minus the skyrmion intensity, plus the background. The calculated stripe phase intensity subtracts the overlapping skyrmion signal, which includes the ring-like background, then reintroduces the background to allow accurate comparisons to the skyrmion intensity. Tracing the feature associated with the skyrmion phase confirms its stability is robust over a wide range of applied fields, between $\mu_{0}H=$~185~mT and $\mu_{0}H=$~-30~mT; temperature scans performed at $\mu_{0}H=$~0~mT once established at 300~K; also confirms a temperature stability from 10~K to $>$320~K. Passing through remanence the intensity of the skyrmion phase decreases, coinciding with the increase of the stripe phase intensity.  The initial increase in intensity of the skyrmion phase closely tracks the increasing intensity of the ring-like background. As noted above, the ring feature corresponds to phases with a common periodicity but no orientation. A likely origin of this feature is therefore the nucleation of labyrinthine domains, perhaps in regions between skyrmion domains, or in regions of skyrmion-lattice defects.

Performing similar analysis on the [(Fe(3.6~\AA)~/~Gd(3.8~\AA)]$\times$120 sample (Figure~\ref{fig:Analysis2}b) identifies similar qualitative trends for the skyrmion lattice. However, in this sample no co-existing stripe phase was seen as we approached saturation (Supplemental Fig.~S4). A slight change in compositions is expected to have a large effect on the skrymion size and correspondingly the intensity of the diffraction pattern of the skyrmion lattice.

\begin{figure*}[ht]
    \centering
    \includegraphics[scale = 0.2]{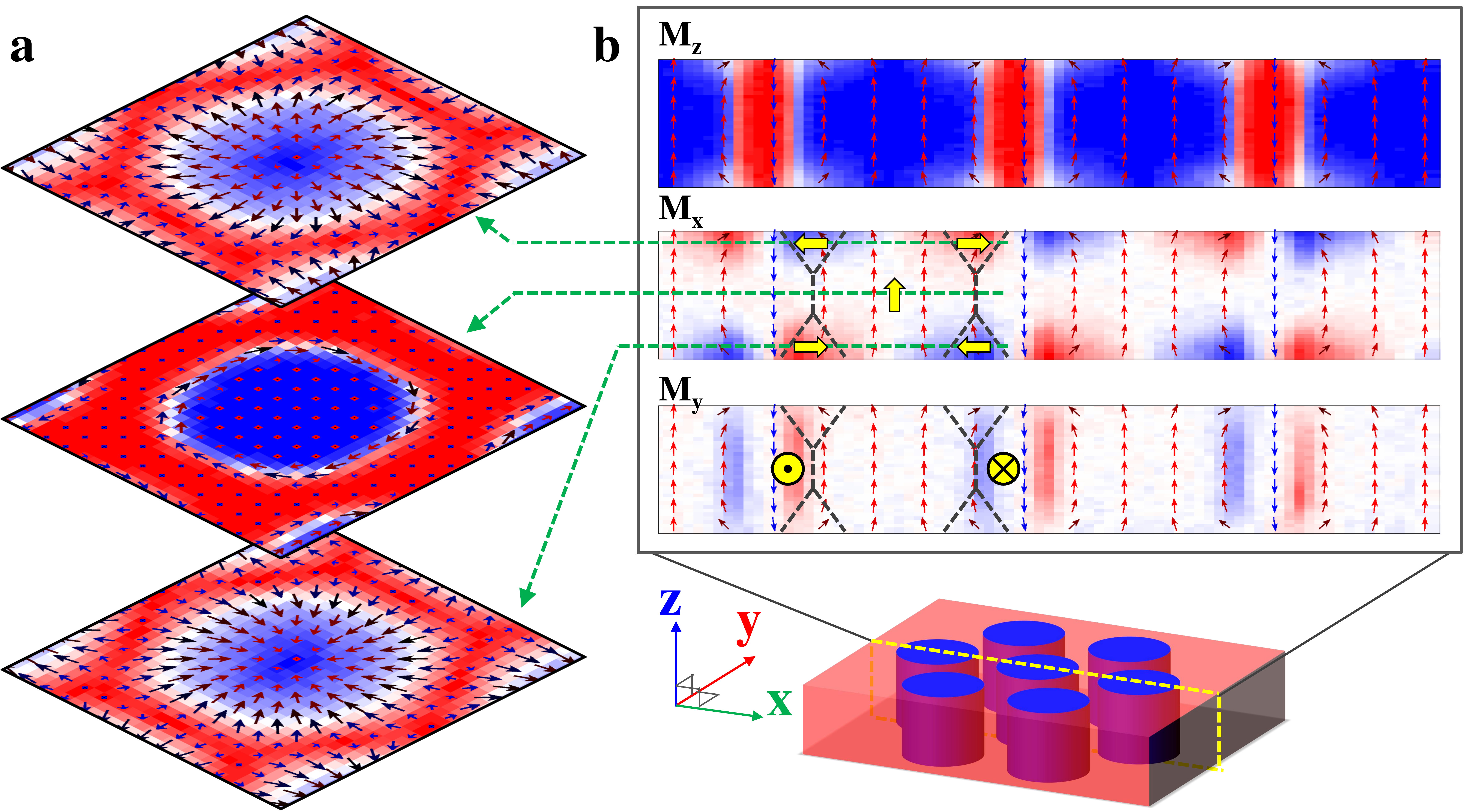}
    \caption{Micromagnetic simulations of the proposed flux closure skyrmion structure confirm its stability at zero field and a simulated temperature of 300~K. The modeled lattice has skyrmions with a core in the $+z$ direction in a matrix oriented in the $-z$ direction. (\textbf{a}) The top-down view shows that the boundary surrounding the core on the top and bottom surfaces are oriented in the radial direction similar to a N\'eel skyrmion, facing away from the core on the top surface and towards the core on the bottom, following the dipole fields. At the equatorial belt of the skyrmion the moments are oriented along the azimuthal direction, similar to a Bloch structure. (\textbf{b}) The cross sectional view of the skyrmion M$_z$ (top) shows the orientation of the core in the +z direction and -z direction of the matrix. The M$_x$ component (middle) confirms the flux closure structure, with the moments on the top and bottom surfaces oriented to follow the dipolar fields. The M$_y$ component (bottom) shows the moments at the barrel of the skyrmion are oriented out of the plane of the page, in opposite directions, due to the bisection of the azimuthally oriented boundary.}
    \label{fig:OOMMF}
\end{figure*}
One key result of the experimentally-observed hexagonal ordering is that these skyrmions possess a mutual repulsive interaction. For Bloch-type skyrmions with random chiralities, and azimuthal-only character, the interaction is not strictly repulsive and skyrmions with the opposite chiralities can condense to form biskyrmions which would not form hexagonal arrays\cite{Lee.2016}. We propose that distortions of the skyrmion structure, in the form of surface-bound flux-closure domains,\cite{Montoya.2017-1} result in a feature that supports the mutual repulsion required to promote lattice ordering. Specifically, in the flux-closure structure, small in-plane domains form at the domain wall/film surface boundary to reduce the stray magnetic fields, shown in Figure~\ref{fig:OOMMF}. The orientation of the domains follow the dipole fields; for a skyrmion with an upward facing core, for example, the dipole fields orient radially outward (away from the core) on the top surface, and radially inward (toward the core) on the bottom surface. This proposed structure is similar to a N\'eel skyrmion structure, but with opposite chiralities on the top and bottom, giving it a net topological charge of zero. However, Lorentz transmission electron microscopy (TEM) results have previously shown that skyrmions in these systems are Bloch-type,\cite{Montoya.2017-2} with moments oriented in the azimuthal direction, seemingly at odds with the flux-closure structure. In the flux-closure structure, azimuthally-oriented moments would be located at the barrel of the skyrmion between the flux-closure domains; TEM measurements would be predominantly sensitive to these azimuthal moments, while the radial moments on the top and bottom would contribute oppositely to the image and thus would tend to cancel out in the transmission geometry, leaving only the Bloch walls. This new hybrid skyrmion construction would possess a net topological charge of unity, with no net contribution from the N\'eel-type flux-closure caps, and unity from the Bloch-type azimuthally wound moments along the barrel.

To investigate the stability of the described flux closure skyrmion structure inferred from our data, micromagnetic simulations performed using OOMMF\cite{donahue1999oommf} shown in  Figure~\ref{fig:OOMMF}. Indeed, the proposed structure was confirmed to be stable at zero field and at simulated temperatures of 300 K. The top-down view of the skyrmions, Figure~\ref{fig:OOMMF}a shows the moments oriented radially on the top and bottom surfaces, with opposite directions relative to the core, following the dipole fields. At the equatorial belt of the skyrmion, the moments are aligned in the azimuthal direction, consistent with the proposed structure and the TEM results. The cross-section of the skyrmions, Figure~\ref{fig:OOMMF}b shows that the magnetization follows the proposed flux closure structure, with small surface domains oriented in-plane away from the skyrmion core on the top, and towards the core on the bottom. The magnetization along the y-direction bisects the azimuthal ring, and thus is into the plane on the right, and out of the plane on the left (for the demonstrated counter-clockwise chirality). This specific structure is crucial for the formation of the hexagonal ordering of the skyrmion lattice and has been observed recently in Pt/Co/Ir and Pt/Co/AlO$_x$ trilayers\cite{Legrand.2018}. These simulations did not include a random anisotropy, and hence demonstrate stability is possible with only dipolar considerations, consistent with previous works\cite{buttner2018theory}. 

\section*{Conclusion}

We have experimentally demonstrated an approach to generate lattices of hybrid N\'eel-Bloch skyrmions with long-range order, stable at room temperature and zero applied magnetic field, in amorphous multilayer thin films of Fe/Gd. These unique structures were achieved by saturating the film with a tilted magnetic field, facilitating the construction of an ordered stripe phase. This field sequence gives rise to a well ordered state of the skyrmion lattice over a wide range of positive and negative fields ($\approx200$~mT), previously unreported, along with a broad range of temperatures (10 K to $>$320 K). This range of stability is crucial for the realization of devices stable against stray fields. Additionally, these hybrid N\'eel-Bloch skyrmions lack the in-plane chiral symmetry breaking present in traditional skyrmions stabilized by DM interactions. As a result, both chiralites can exist in these films and can be uniquely controlled, providing additional avenues of manipulation for recording tertiary data-bits, for example. Alternatively, long-range chirality control may be realized within the system by adding additional exchange-coupled layers which possess traditional DM interactions. The ability to stabilize skyrmions at ambient conditions as well as a broad range of fields makes Fe/Gd multilayered thin films ideal candidates for integration into spintronic devices.

\section*{Acknowledgements}
The authors wish to acknowledge the contributions from John Smith (ORNL), Thomas Farmer (ISIS), and Ken Littrell (ORNL). Work at UCSD was supported by the research programs of the U.S. Department of Energy (DOE), Office of Basic Energy Sciences (Award No. DE-SC0003678). Work at the Advanced Light Source, Lawrence Berkeley National Lab (LBNL) was supported by the Director, Office of Science, BES, of the DOE (Contract No. DE-AC02- 05CH11231). Mi-Young Im acknowledges support by the National Research Foundation (NRF) of Korea funded by the Ministry of Education, Science and ICT (2018K1A4A3A03075584, 2016M3D1A1027831, 2017R1A4A1015323) and by the DGIST R\&D program of the Ministry of Science, ICT and future Planning (18-BT-02). Access to VSANS was provided by the Center for High Resolution Neutron Scattering, a partnership between the National Institute of Standards and Technology and the National Science Foundation under Agreement No. DMR-1508249. These results use the resources at the High Flux Isotope Reactor, a DOE Office of Science User Facility operated by the Oak Ridge National Laboratory.


%

\end{document}